\documentclass[preprint,10pt,authoryear,5p,twocolumn]{elsarticle}

\usepackage{amssymb}
\usepackage{amsmath}
\usepackage{url}
\usepackage{multirow}

\usepackage{hyperref}

\usepackage{ulem, xcolor}

\begin{document}

\begin{frontmatter}

\title{Neural networks in the search for fast radio bursts with~RATAN-600}

\author[1]{D.\,O.~Kudryavtsev\corref{cor1}}
 \ead{dkudr@sao.ru}
\author[1]{S.\,A.~Trushkin}
\author[1]{P.\,G.~Tsybulev}
\author[1,2]{V.\,A.~Stolyarov}
\cortext[cor1]{Corresponding author}

\affiliation[1]{
organization={Special Astrophysical Observatory of the Russian Academy of Sciences},
city={Nizhny~Arkhyz},
postcode={369167},
country={Russia}}

\affiliation[2]{
organization={Astrophysics Group, Cavendish Laboratory, University of Cambridge},
city={Cambridge},
postcode={CB3\,0HE},
country={UK}}

\begin{abstract}
We present a technique to search for fast radio bursts in records obtained with broadband radiometers having few radio channels. The technique is applied to the RATAN-600 surveys carried out at its Western Sector since the year 2017. A 1D convolutional neural network for multichannel time series classification is developed based on the EfficientNet family of models. The procedure to generate synthetic FRB signals needed for the training dataset is described. We implement a two-stage cascade scheme to effectively suppress the rate of false positive detections. Evaluation of the trained model is provided based on the synthetic events and the giant pulse of the Crab Pulsar. 
\end{abstract}


\end{frontmatter}

\section{Introduction}

Fast radio bursts (FRBs) are short-timescale radio-wave transient events caused by some high-energy astrophysical processes not yet understood. FRBs are commonly detected by the characteristic time delay of pulses arriving in different radio frequency bands (channels). This delay is caused by the interstellar and/or intergalactic dispersion and provides the means to effectively distinguish a radio pulse of a cosmic source from the radio frequency interference (RFI) of terrestrial or near-Earth origin. 

The search for fast radio bursts (FRBs) is complicated by several circumstances: despite a large number of expected events, their detections are extremely rare because of the short duration of an event (from a fraction of a millisecond to tens of milliseconds), the small field of view at any given time, and a large number of RFI signals in the daytime. This paper addresses the problem of FRB detection in the data obtained on broadband radiometers with few radio channels, particularly those which are used on the radio telescope \mbox{RATAN-600} of the Special Astrophysical Observatory. We develop a 1D convolutional neural network adapted to search for FRBs in the few-channel time series and consider a cascade scheme to suppress the false positive detections, which can be abundant when a neural network operates on a full range of original data without preliminary algorithmic selection of possible candidates.

Since 2017, RATAN-600 has been regularly conducting, on its Western Sector 
with an antenna effective area of 1200~m$^2$, blind surveys of different sky areas. In eight surveys (strips of the sky with a length of $360^\circ$ and a width of about $0.\!\!^{\circ}6$) at fixed declinations from $5^{\circ}$ to $33^{\circ}$, performed with a four-beam four-channel radiometer at 4.7~GHz with a total bandwidth of 600~MHz and a sensitivity in each beam of about 10~mJy (FRB fluences $>1$~Jy\,ms), extensive  observed data have been accumulated for the total sky area of about 1700~square degrees with a sampling time interval of 0.245~ms.

\begin{figure}[t]
\centering
\includegraphics[width=\linewidth]{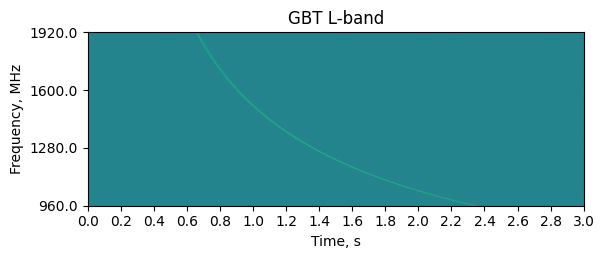}
\includegraphics[width=\linewidth]{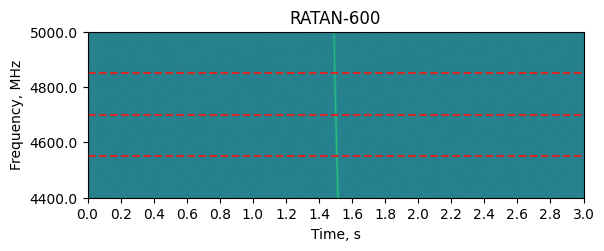}
\caption{A synthetic FRB signal ($w=5$~ms, ${\rm DM}=500$~cm$^{-3}$\,pc) in the time--frequency domain for different frequency ranges: L-band (1--2~GHz, top panel) and RATAN-600 radiometers (4.4--5~GHz, bottom panel). The horizontal red lines designate the boundaries of the RATAN-600 four subchannels.}
\label{fig:profiles} 
\end{figure}

Figure~\ref{fig:profiles} depicts a typical FRB as it can be observed in the time--frequency domain. The top panel shows the FRB at lower radio frequencies, where these events are most frequently detected, while the bottom panel presents the same pulse as it could be obtained on RATAN-600. We can note here two major differences: (1)~the overall duration of the pulse is smaller when observed at higher frequencies; (2)~instead of several thousands of narrow-band radio subchannels, at \mbox{RATAN-600} we currently have four broadband ones. 

\begin{figure}[t]
\includegraphics[width=0.49\linewidth]{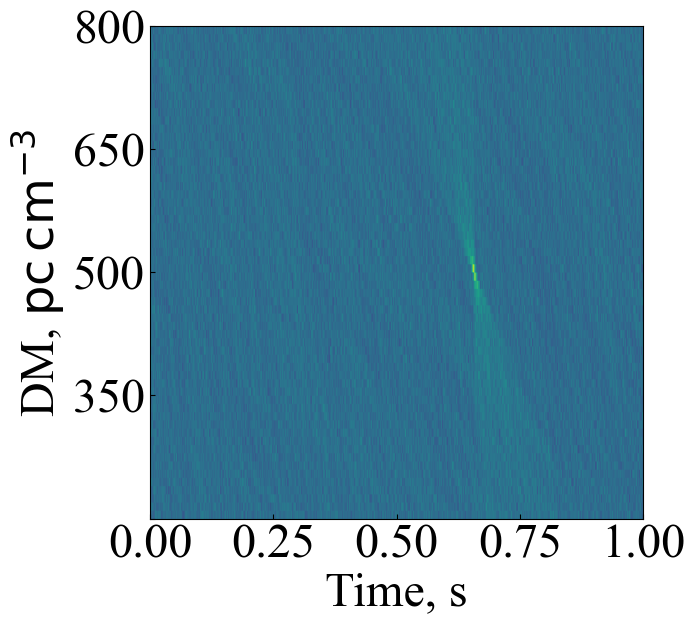}
\includegraphics[width=0.49\linewidth]{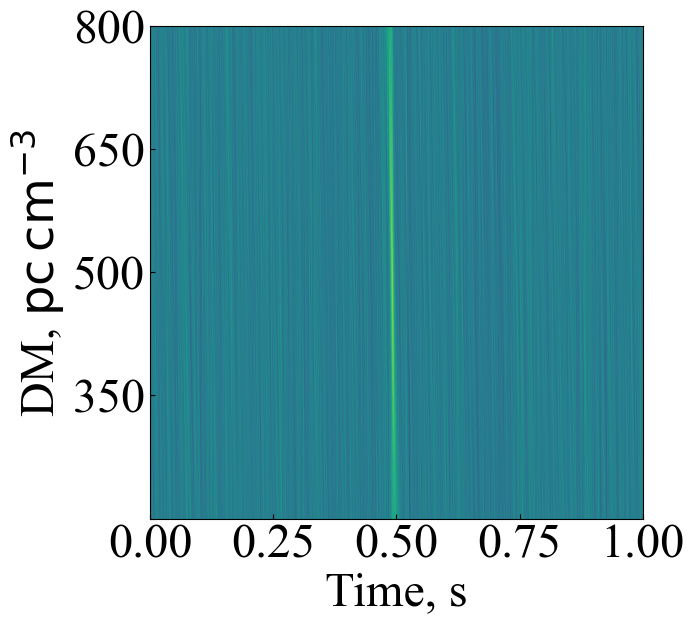}
\caption{Comparison of synthetic DM--time diagrams in the case of a large number of channels (left panel) and the four channels of RATAN-600 (right panel). The small number of channels produces poor localization of pulse parameters.
}
\label{fig:dmt_compar}
\end{figure}

The observed time lags are described numerically by a special characteristic, dispersion measure $\rm DM$, expressed in cm$^{-3}$\,pc, as physically it is the column density of free electrons along the line of sight. The classical methods based on the dedispersion and summing of a possible signal with different values of $\rm DM$ and constructing  in such a way a $\rm DM$--time diagram will not be efficient in the case of a small number of radio channels. As an example, Fig.~\ref{fig:dmt_compar} shows the $\rm DM$--time diagrams for the same synthetic pulse in the cases of the large (4000) and small (4) number of channels. The localization of the event along the DM axis is poor in the right panel. Not only would it be harder to measure the DM of the pulse but the great uncertainty in the DM would make it difficult to distinguish between FRB pulses and RFI signals.

There exist methods to search for possible FRB pulses using machine learning models, namely convolutional neural networks (CNNs). These can take as the input data either the original ``time~--~radio frequency'' records or images in the $\rm DM$--time domain or both, but in any case the models are meant to process 2D images, which again is not the case of RATAN-600 data. Here the input ``time~--~radio frequency'' data are arrays with only four elements along one of the axes. Such data are inherently better described as a combination of several vectors rather than a 2D image. With few ``pixels'' along one of the axes, the receptive field of neurons in the deeper layers of a 2D-convolutional network will be the same along that axis, which ruins the whole point of using the second dimension. Moreover, a 2D convolution with a $k\times k$ kernel applied to an image with a size of $n\times n$ have a computational complexity $O(n^2 k^2)$, while for a corresponding 1D convolution with the same $n$ and $k$ this is $O(n k)$ (e.g., \citealt{KIRANYAZ2021107398}). This gives us at least a factor of $k$ advantage in the case of multichannel 1D convolutions, and the typical kernel size in modern CNNs is $k=3$. Images in the \mbox{DM--time} domain should have been processed with 2D CNNs, but we did not use this representation for the already mentioned reason of poor DM localization (Fig.~\ref{fig:dmt_compar}).

Thus, the configuration of the receiving equipment dictates the need for specific data processing techniques, and this paper is dedicated to applying 1D convolutional neural networks to search for FRBs in RATAN-600 observations. Although designed for this specific task, the developed neural network can also be used for general-purpose classification of multichannel time series. 

The PyTorch-based code of the neural network architecture is available on GitHub\footnote{\url{https://github.com/DKudryavtsev/EfficientNet1D_FRB}} along with the code to generate synthetic FRB events, used for the training sample.   

\section{Neural Network Architecture}
\subsection{Previous works}

Successful detection of FRBs using the machine learning approach has already been made previously, e.g., on the Green Bank Telescope (GBT, \citealt{2020MNRAS.497..352A}) and Arecibo \citep{2022MNRAS.509.1929P}. In their paper, \cite{2020MNRAS.497.1661A} generally describe the conception behind the use of deep learning methods, specifically the convolutional neural networks (CNNs), for the search of FRB events. They also list the models implemented earlier, particularly the multi-input CNN with two convolutional and two pooling layers of \cite{2018AJ....156..256C} and the \mbox{17-layer} ResNet-based architecture of \cite{2018ApJ...866..149Z}, where the input data were the initial, dispersed, frequency--time spectrograms. As well, \cite{2020MNRAS.497.1661A} develop their own {\tt FETCH} classifier for fast transient classification, which contains eleven deep-learning models based on various architectures. The input data for {\tt FETCH} are frequency--time and DM--time images.

The principal distinction of RATAN-600 observing data from the data used in the above mentioned approaches is that the latter consist of time series recorded in many narrow radio channels (4096 channels for the GBT) that makes it possible to represent them as a 2D frequency--time image with individual radio channels unfolded along one of the image axes. This image could then be fed into common convolutional neural networks developed, e.g., for classification of objects in usual photos.
In the case of \mbox{RATAN-600} observations, we currently have only four channels, each collecting the radio flux in a broad 150-MHz bandwidth. 
As is already mentioned in the Introduction, this format of data is far from the 2D rectangular image representation and can be better described as a 1D array with four channels. To classify this type of data, it is more convenient to use 1D convolutions (e.g., \citealt{KIRANYAZ2021107398}), which have a lower number of trainable parameters and lower computational cost while performing the task with the same accuracy as their 2D analogs.

\cite{IsmailFawaz2018DeepLF} give a review of a number of 1D CNN architectures for time series classification and compare their performance. Two of the considered models demonstrate the best metrics: the Fully Convolutional Network (FCN) and the Residual Network (ResNet), both suggested by \cite{7966039} based on their 2D predecessors introduced by \cite{7298965} and \cite{7780459}. 

For our project we adopted the newer EfficientNet family of models \citep{pmlr-v97-tan19a, pmlr-v139-tan21a} instead of the FCN or ResNet. A comparison of the FCN and EfficientNet performances, in favor of the latter model, is given further in Section~\ref{sec:eff_vs_fcn}. As for the ResNet architecture, we did not try to implement it in this paper for the following reasons.
\begin{itemize}
\item 
EfficientNet follows the same strategy as ResNet: increasing the depth of the network to increase the number of parameters and extract finer features. Both architectures use residual connections to overcome the problem of vanishing gradients. In this sense, \mbox{EfficientNet} can be regarded as an advanced version of ResNet.
\item 
EfficientNet is based on mobile inverted bottleneck convolution (MBConv) blocks \citep{8578572} that are specifically designed to reduce the computational cost while maintaining high accuracy.
\item 
EfficientNet addresses the overfitting by implementing the Stochastic Depth \citep{10.1007/978-3-319-46493-0_39} and Dropout \citep{JMLR:v15:srivastava14a} techniques.
\item 
In traditional (2D) image classification problems, \mbox{EfficientNets} outperform ResNets, whereas being smaller and working faster.
\end{itemize}

We have constructed our own PyTorch-based 1D version of EfficientNet by modifying and scaling the \mbox{EfficientNetV2} architecture \citep{pmlr-v139-tan21a}.

\subsection{EfficientNet1d-XS}

Our modifications applied to the EfficientNetV2-S baseline architecture \citep{pmlr-v139-tan21a} are as follows.
\begin{itemize}
\item Two-dimensional convolutions have been replaced with 1D convolutions.
\item \citep{pmlr-v97-tan19a} give a recipe for compound scaling of the EfficientNet baseline, where the scaling should be applied jointly to the width (number of channels) and depth (number of layers) of the baseline neural network as well as to the resolution of the input image. In our case we cannot scale the resolution of the input time series because high temporal resolution is critical for FRB detection. Having in mind that processing 1D data is much simpler than processing a 2D image, we have just scaled the width and depth of the 2D baseline model (EfficientNetV2-S) proportionally to the minimum, so that to reduce, e.g., two layers in a block to one layer, four layers to two layers, and so on. Our network is thus even smaller than the baseline, therefore we have added the suffix ``XS'' (extra small) to its name.
\item Residual connections in MBConv blocks are just the input of a block, as usual; but unlike in the original implementation, we have also added residual connections in the blocks where the input and output tensors have different shapes (different number of channels or a reduced size in the output layer due to a stride~$>1$). In these cases the residual connections are convolutions with a kernel size $k=1$ and/or an appropriate stride, followed by batch normalization (BN, \citealt{10.5555/3045118.3045167}). This is the same as the residual connections in the original ResNet architecture.
\item The reduction ratio in the Squeeze-and-Excitation (SE) block \citep{8578843} is taken to be 1/16, according to the recommendation of the SE developers. The SE block is a part of the MBConv block (see below).
\item We have replaced the Rectified Linear Unit (ReLU) activations with the Gaussian Error Linear Units (GELU, \citealt{hendrycks2023gaussian}), which demonstrate good performance in a number of deep learning problems.
\end{itemize}

We did not experiment with network hyperparameters, so we cannot state that this architecture is fully optimized for our particular problem or for time series classification in general; nevertheless, it allowed us to obtain very accurate results in a reasonable amount of training and inference time. The capacity of the network is expected to be sufficient enough to solve a variety of problems related to 1D data analysis by deep learning methods.

The hyperparameters of EfficientNet1d-XS are listed in Table~\ref{tab:effnet}. Note that strides~$=2$ are applied only to the first layer in a corresponding block (\# layers), for other layers in the block strides~$=1$. The input data in our case is a tensor of size [${\rm B} \times {\rm C} \times {\rm L}$], where ${\rm B}$ is the batch size, $\rm{C}=4$ is the number of channels, $\rm L=4080$ is the time series length ($\sim\!1$~s at the radiometer temporal resolution of 0.245~ms). The final fully connected (FC) layer has one neuron as its output, which should signal the presence of an FRB event in the record. The input and output layers may be easily modified for other data shapes. 

\begin{table}[t]
\small
\centering
\caption{EfficientNet1d-XS architecture}
\label{tab:effnet}
\begin{tabular}{ccccc}
\hline
Operator & Kernel & Stride & \# channels & \# layers \\
\hline
Conv, BN, GELU  & 3  & 2  &  12 & 1 \\
Fused MBConv1   & 3  & 1  &  12 & 1 \\
Fused MBConv2   & 3  & 2  &  24 & 2 \\
Fused MBConv2   & 3  & 2  &  32 & 2 \\
MBConv2         & 3  & 2  &  64 & 3 \\
MBConv3         & 3  & 1  &  80 & 4 \\
MBConv3         & 3  & 2  & 128 & 7 \\
Conv            & 1  & 1  & 640 & 1 \\
GlobAverPool    & -- & -- &  -- & 1 \\
Dropout 0.2, FC & -- & -- &  -- & 1 \\
\hline
\end{tabular}
\end{table}

\begin{figure}[t]
\centering
\includegraphics[width=\columnwidth]{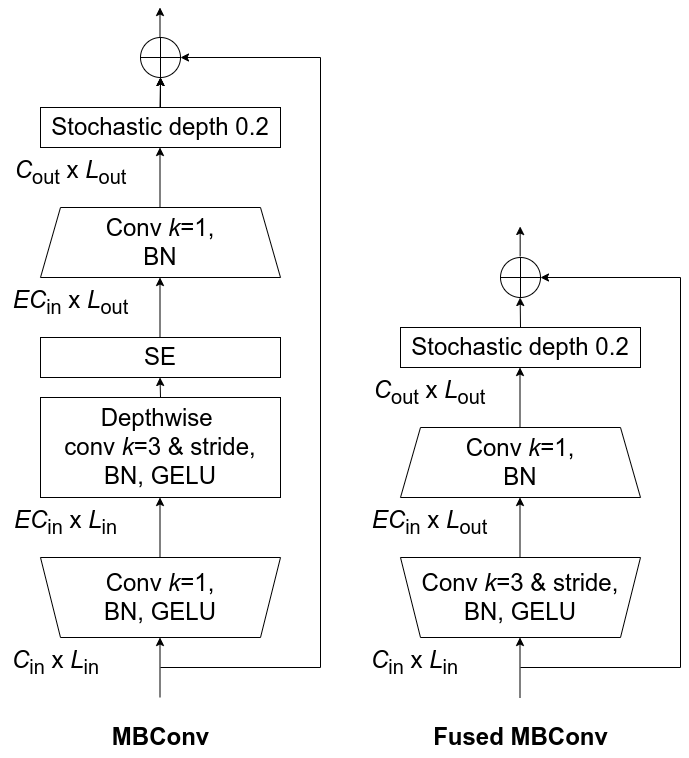}
\caption{MBConv and Fused MBConv blocks. The transformations of the processed tensor shape are indicated below the blocks. $C_{\rm in}$ and $C_{\rm out}$ is the input and output number of channels, $L_{\rm in}$ and  $L_{\rm out}$ is the input and output length of a processed tensor, and $E$ is the expansion factor.}
\label{fig:mbconv}
\end{figure}

The structure of the MBConv (mobile inverted bottleneck convolution, \citealt{8578572}) and Fused MBConv\footnote{https://research.google/blog/efficientnet-edgetpu-creating-accelerator-optimized-neural-networks-with-automl/} blocks is analogous to those in the original EfficientNet architecture and is shown in Fig.~\ref{fig:mbconv}. 

The main specificity of MBConv block operation is the widening of the number of channels in the hidden layer with following depthwise (grouped) convolution \citep{NIPS2012_c399862d}, which greatly reduces the number of parameters while allows a network to learn finer features. In Fused MBConv blocks the depthwise convolution is omitted, which increases training speed if applied in the early stages \citep{pmlr-v139-tan21a}. The expansion factor $E$ is indicated in Table~\ref{tab:effnet} in the names of MBConv blocks, e.g. MBConv2 indicates $E=2$. 

The SE block (Table~\ref{tab:se}, \citealt{8578843}) assigns trainable weight factors for channels in the MBConv hidden layer, thus allowing for their importance in the final $k=1$ convolution. The SE block itself is Global Average Pooling \citep{lin2014network} followed by two fully connected (FC) layers and a Sigmoid activation in the output. The reduced dimension between the FC layers is taken to be 1/16 of the number of channels in the MBConv hidden layer. 

\begin{table}[t]
\small
\centering
\caption{Squeeze-and-excitation block}
\label{tab:se}
\begin{tabular}{ll}
\hline
Operator & Shape \\
\hline
GlobAverPool & \# channels \\
FC, GELU     & \# channels / 16\\
FC, Sigmoid  & \# channels\\
\hline
\end{tabular}
\end{table}

Finally, Stochastic Depth \citep{10.1007/978-3-319-46493-0_39} is implemented as the output of the block. This layer randomly turns off an entire MBConv block during training by zeroing its output, leaving only the residual connection to pass the information. This is a part of regularization technique along with Dropout \citep{JMLR:v15:srivastava14a} before the last FC layer (see Table~\ref{tab:effnet}). For both operators we took the probability of dropping $p=0.2$.

The total number of trainable parameters for our configuration (four input channels, one output neuron) is 1\,125\,669. The relatively low number of parameters and the network architecture allow the inference to be performed using even a modest gaming GPU.

\subsection{Comparison with the FCN}
\label{sec:eff_vs_fcn}

Table~\ref{tab:eff_vs_fcn} gives a comparison of some characteristics of the 1D FCN and EfficientNet architectures. The evaluation technique used here is described in more detail further in Section~\ref{sec:eval}. The FCN hyperparameters were adopted from~\cite{7966039}. The inference time has been estimated using an NVIDIA GeForce RTX~3080 graphics processing unit (GPU). The detection rates are estimated based on synthetic FRB pulses and given for two signal-to-noise ratios: $\rm SNR=3$ and 10. As well, two kinds of pulses were used: (1)~``average'' FRBs with a pulse width $w=5$~ms and a dispersion measure $\rm DM=500~cm^{-3}\,pc$ and (2)~more narrow pulses with $w=2$~ms and a smaller $\rm DM=200~cm^{-3}\,pc$.

\begin{table*}[t]
\small
\centering
\caption{
Comparison of the FCN and EfficientNet characteristics}
\label{tab:eff_vs_fcn}
\begin{tabular}{lcc}
\hline
Characteristic & FCN & EfficientNet \\
\hline
Trainable parameters & 267\,393 & 1\,125\,669 \\
No. of layers & 3 Conv & 19 MBConv \\
Inference time (per instance) & $0.55\pm0.01$ ms & $1.95\pm0.05$ ms\\
False positive rate (test sample) &  0 & $0.0065\pm0.0008$ \\
\hline
FRB detection rates: & & \\
$\rm SNR = 3$, $w = 1$~ms, $\rm DM=200~cm^{-3}\,pc$ & $0.40\pm0.01$ & $0.55\pm0.01$ \\
$\rm SNR = 3$, $w = 5$~ms, $\rm DM=500~cm^{-3}\,pc$ & $0.64\pm0.01$ & $0.977\pm0.004$ \\
$\rm SNR = 10$, $w = 1$~ms, $\rm DM=200~cm^{-3}\,pc$ & $0.86\pm0.01$ & $1$ \\
$\rm SNR = 10$, $w = 5$~ms, $\rm DM=500~cm^{-3}\,pc$ & $0.999\pm0.001$ & $1$ \\
\hline

\hline
\end{tabular}
\end{table*}

Having four times as many parameters as the FCN, the EfficientNet architecture demonstrates a roughly proportional increase in inference time. At the same time, it has a depth of 19 convolutional layers (MBConv blocks) in contrast to the 3 layers in the FCN, which provides the ability to learn finer features, and this is confirmed by the much better detection rates for all kinds of the pulses (see the bottom rows in Table~\ref{tab:eff_vs_fcn}). Although the FCN has not produced false positive results, this is achieved at the cost of worse detection rates for all the variety of synthetic FRBs with different characteristics. Thus, EfficientNet is definitely more suitable for our project, especially in detection of weaker and narrower pulses.

\section{Dataset}

\subsection{Radiometer noise and radio frequency interference}

The dataset consisted of three equal parts of events: real noise of the radiometers, real RFI signals observed, and synthetic FRBs injected into the real noise records. We used time intervals of 1~s to divide the time series into a sequence of separate events.

The radiometer noise records were taken during the RFI-quiescent time period from 11pm to 06am. In this period, there are typically no or just a few RFI signals. These rare signals do not degrade the training performance as their percentage is negligibly small and, anyway, they also must be classified as the absence of an FRB along with the true noise records.

The RFI part is a bit trickier to obtain. The RFI time profiles are very diverse, from short pulse events to the small smooth increase of the radiometer noise, and there is no algorithm able to extract all the variety, especially for the low-amplitude events. The manual labeling would have been incredibly time consuming, and the time profiles are hard to be reliably synthesized. To obtain the RFI subsample, we used a neural network trained with only the noise records and synthetic FRBs. After the training, it happened to be able to detect any pulses different from the noise records, thus providing us a tool to collect the required RFI subsample almost immediately.

In total, we collected nearly 100\,000 RFI and 200\,000~radiometer noise records, the latter to be divided into the subsamples of the actual noise records and the noise records with synthetic FRBs injected.

\subsection{Synthetic FRBs}
\label{sec:synth_frb}

\subsubsection{Time delay}

The synthetic FRBs were generated similarly to \cite{2020MNRAS.497..352A} with slight modifications. The code developed to synthesize FRB events was partially adopted from {\tt injectfrb}.\!\footnote{https://github.com/liamconnor/injectfrb}
The pulse time profile $s(t)$ is initially a Gaussian with varying intrinsic width for different events. The pulse delays in different bands are calculated according to the standard formula (e.g., \citealt{2019fraa.book.....S}):
\begin{equation}
\Delta t = 0.41\times10^4\,{\rm DM}\left(\frac{1}{\nu_1^2}-\frac{1}{\nu_2^2}\right),
\end{equation}
where $\Delta t$ is the time delay (in seconds) between a pair of frequencies $\nu_1<\nu_2$ expressed in MHz, and $\rm DM$ is the dispersion measure in cm$^{-3}$\,pc. 

\subsubsection{Scattering}

The scattering is added to the pulse by convolving its time profile with an exponential function $e^{-t/\tau_\nu}/\tau_\nu$, where $\tau_\nu$ is the scattering time at a frequency $\nu$. We assume the power-law frequency dependence $\tau_\nu\propto\nu^{-4}$. With scattering times of up to 1~s at 1\,GHz, the effect is negligible at our frequencies of about 4.7~GHz, but we have nevertheless included it in the modeling for the general case.

\subsubsection{Dispersion smearing}

The dispersion smearing caused by the pulse delay within the frequency range of a radiometer channel is then applied by the convolution of the time profile with a sequence of $N$ ones, where $N$ can be determined as
\begin{equation}
N = \lfloor t_{\rm DM} / t_{\rm samp}\rceil.
\end{equation}
Here $t_{\rm DM}$ is the time during which the pulse is retained in the channel because of its dispersion, $t_{\rm samp}$ is the radiometer sampling time (temporal resolution), and the brackets~$\lfloor\rceil$ designate rounding to an integer value. The $t_{\rm DM}$ value can be calculated from Eq.~(1) taking the channel boundary frequencies.

\subsubsection{Per-channel amplitudes}

Variation of time profile width across different channels is naturally calculated in the convolutions, but we need to adjust the model pulse amplitudes in the channels accordingly. To this end, the calibration of the amplitude in a channel is performed by
\begin{equation}
s_\nu(t) = \frac{s_\nu(t)}{\max s_\nu(t)} \frac{w_{\rm int}}{\sqrt{w_{\rm int}^2+t_{\rm DM}^2+\tau_\nu^2}}\,A_{\rm int}, 
\end{equation}
where $w_{\rm int}$ and $A_{\rm int}$ are the intrinsic pulse width and amplitude.

In the calculations, we can additionally adjust the per-channel amplitudes for the intrinsic spectral index $\alpha$ of the pulse: 
\begin{equation}
s_\nu(t) =  s_\nu(t) (\nu/\nu_{\rm ref})^\alpha, 
\end{equation}
where $\nu_{\rm ref}$ is the reference (central) frequency of the entire observed frequency band. 

As well, a scintillation/nulling pattern can be applied using a trigonometric function:
\begin{equation}
\begin{split}
s_\nu(t) & = k\,s_\nu(t) \\
k(\nu) & = \cos[2\pi n(\nu/\nu_{\rm ref})^2 + \phi],
\end{split}
\end{equation}
where $k$ is the signal suppression in a particular channel. The number of cycles $n$ in our modeling is drawn from a distribution $e^{U(a, b)}$, where $U(a, b)$ is the uniform distribution. Values $n<1$ are considered as the absence of scintillations. The phase $\phi$ is drawn from the uniform distribution $U(0, 1)$. Allowing for scintillations in the training is critically important in our case because of the small number of radiometer channels.

\subsubsection{Parameter ranges}

The FRB parameters used in the generation of the synthetic FRB sample are drawn from the distributions summarized in Table~\ref{tab:ranges}. The intrinsic amplitudes $A_{\rm int}$ are referred to a single radio channel, expressed in the units of radiometer noise variance $\sigma_{\rm noise}$, and have an additional restriction \mbox{$A_{\rm int} > 2\sigma_{\rm noise}$} to avoid too faint FRB pulses in the training. They further can be additionally reduced by applying the scintillation pattern. The adopted distribution and the condition $n>1$ for the scintillation pattern allow us to obtain a sample where approximately 77\%\ of the events demonstrate scintillation/nulling effects, while the rest are pure FRB signals. Some events can even be extreme cases where the signal is nulled in all but one channel. 

\begin{table}[t]
\small
\centering
\caption{
Distributions of FRB parameters in the synthetic FRB sample
}
\label{tab:ranges}
\begin{tabular}{lclc}
\hline
\multirow{2}{*}{Parameter} & \multirow{2}{*}{Units} & \multirow{2}{*}{Distibution} & Distribution \\
& & & range \\
\hline
Amplitude $A_{\rm int}$ & $\sigma_{\rm noise}$ & Log-normal & $\mu=3$, $\sigma=1$ \\
Width $w_{\rm int}$ & ms & Uniform & [$0.02$, $50$]\\
$\rm DM$ & cm$^{-3}$\,pc & Uniform & [$30$, $5000$] \\
Scattering $\tau_{\rm 1GHz}$ & s & Uniform & [$0$, $1$]\\
Sp. index $\alpha$ & -- & Uniform & [$-4$, $4$] \\
Scintillation $n$ & -- & $\exp({\rm Uniform})$ & [$\ln(0.5)$, $\ln(10)$] \\
Scintillation $\phi$ & -- & Uniform & [$0$, $1$] \\
\hline
\end{tabular}
\end{table}

\begin{figure*}[t]
\centering
\includegraphics[width=\linewidth]{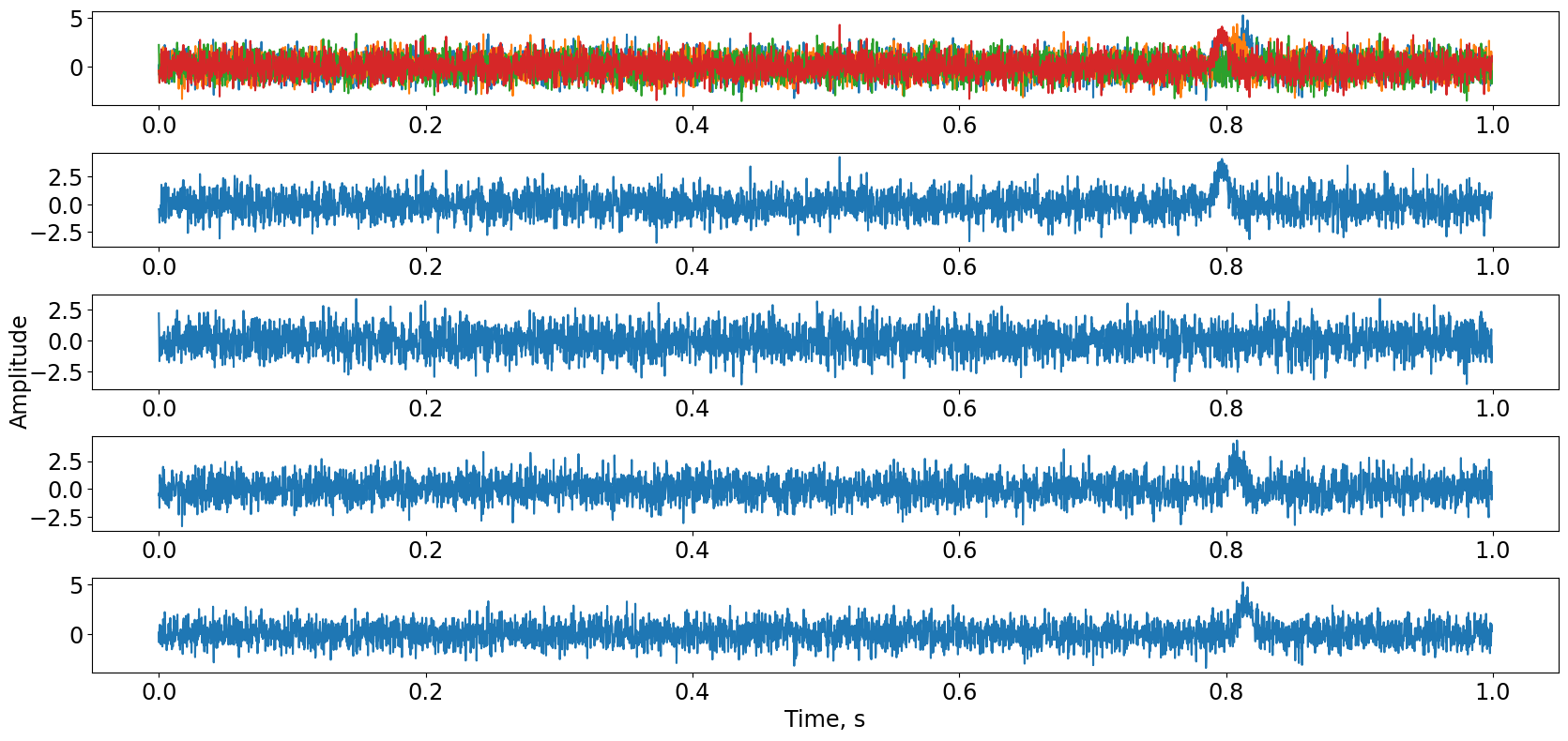}
\caption{An example of a synthetic FRB event: $A_{\rm int}=3\,\sigma_{\rm noise}$, ${\rm DM}=500$~cm$^{-3}$\,pc, $w_{\rm int}=5$~ms. The four lower panels demonstrate the records in the four radiometer channels, in the top panel the records are combined. The nulling effect is seen in the second channel.}
\label{fig:synth_frb}
\end{figure*}

After the modeling, synthetic signals are injected into individual radiometer noise records to form the FRB part of the training sample. The position of the signals are of course varied within the records. In Fig.~\ref{fig:synth_frb} we show an example of a synthetic FRB.

\section{Training}

\subsection{General Notes}

In our case the input data are four time series, each 4080~elements long, which at the radiometer temporal resolution of about 245~{\textmu}s have a time duration of about 1~s. Each channel is scaled to the noise variance $\sigma_{\rm noise}=1$. In the neural network we interpret this data as a \mbox{4-channel} 1D image. The task is simple binary classification: classifying the image either as an FRB event or the absence of it, i.e., the model does not distinguish between pure radiometer noise records and RFI. The output layer is therefore a single neuron signaling the presence of an event, and the loss function is binary cross-entropy.

We used the standard division of the dataset into the training/validation/test samples: 64\%, 16\%, and 20\%, respectively, i.e., the proportion of the validation subsample in the total training and validation sample is also 20\% ($16/(64+16)$): the same as the ratio of the test subsample to the total dataset size. The validation sample was used to track overfitting and choose the optimal checkpoint. In dividing the RFI signals between the samples, we took different hours for different samples to prevent data leakage, as RFI is often detected at all the radiometers simultaneously.

Concerning the proportions between the train\-ing/test/vali\-da\-tion samples, there is no exact rule on this matter. The general requirements are that the diversity of the instances should be represented in all the samples, the training sample should be reasonably large, and both the validation and test samples should be of a size sufficient to make statistical estimations. With the size of our dataset of approximately 300\,000 instances, we have about 48\,000 instances in the smallest validation sample, or 16\,000 events of each type (FRB/noise/RFI), which we consider as a sufficient amount. The diversity is guaranteed by random generation and by shuffling the events during dataset division. 

Data augmentation was implemented by the flip and channel permutation in the noise and RFI records. Generation of an FRB event as described in Section~\ref{sec:synth_frb} is not computationally expensive; this fact allowed us to use a special kind of augmentation: continuous generation of FRB events directly in the training process so that they are never exactly repeated even in different training epochs. The synthetic FRBs in the validation and test samples, though, were generated as usually beforehand for the consistency of evaluation.

The training was performed using the PyTorch \citep{NEURIPS2019_9015} and PyTorch Lightning \citep{2020zndo...3828935F} frameworks.

\subsection{The Cascade Scheme}

With the four radiometers and the 1-s duration of the time series, we have 345\,600~instances per day to be analyzed. In this situation even a model with the best metrics of classification will produce quite a large number of false positive (FP) detections. Along with this, preliminary selection of candidates by usual methods may be ineffective in the case of the small number of channels (Fig.~\ref{fig:dmt_compar} in Introduction). 

To deal with this problem, we use the cascade scheme with two models, which can be dubbed ``expert~1'' and  ``expert~2.\!'' Expert~2 performs the same task of classification but is trained not on the all variety of RFI events but only on those that are FP detections from expert~1. The results is that the overall FP rate is reduced to a product of the FP rates of the two models. The true positive (TP) detection rate is reduced as well, but with good classification metrics the efficiency of the cascade does not reduce much.

We used the same neural network architecture (EfficientNet1d-XS) for both models, but the training was performed differently (Table~\ref{tab:training}). Note though that this inequality in the training procedure is simply a result of a number of training runs to achieve the best evaluation metrics and should not be considered as the only way to train the best model. 

We also found that choosing the optimal checkpoint by validation on the narrow FRB pulses with low $\rm DM$ in the training of expert~1 improves the detection of the more difficult, for the network, part of FRB pulses whose time profiles are harder to distinguish from the RFI signals (the last row of Table~\ref{tab:training}).

\begin{table*}[t]
\small
\centering
\caption{Training of the two neural networks in the cascade scheme}
\label{tab:training}
\begin{tabular}{ll}
\hline
Expert~1 & Expert~2 \\
\hline
Full RFI sample & RFI signals missed by expert 1 \\
Time flip (for RFI and noise) & Time flip (for RFI and noise) \\
Channel permutation (for RFI and noise) & Channel permutation (for RFI and noise) \\
Continuous FRB generation in the training subsample & Fixed FRB samples \\
Validation on narrow low-$\rm DM$ FRB pulses & Validation on all FRB pulses \\
\hline
\end{tabular}
\end{table*}

\subsection{Training process}

\begin{figure}[t]
\centering
\includegraphics[width=0.9\linewidth]{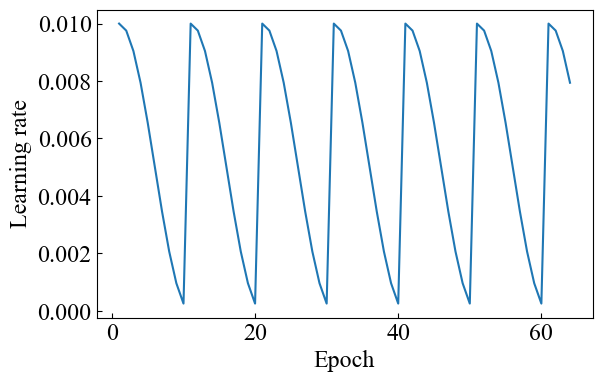}
\caption{
Variation of the learning rate at the beginning of each epoch according to cosine annealing with warm restarts.
}
\label{fig:annealing}
\end{figure}

The models were trained with an NVIDIA GeForce RTX~3080 graphics card (8700~CUDA cores, 10~GB GDDR6X) using a batch size of 256~records. The initial learning rate was estimated automatically using PyTorch Lightning Tuner and amounted to $\sim$0.01 and $\sim$0.003 for expert~1 and expert~2, respectively. We additionally adjusted its value using cosine annealing with warm restarts after 10 epochs (Fig.~\ref{fig:annealing}).
Within the epochs the learning rate was regulated by the Adamax optimizer. We set a sufficient number of epochs with an early stopping in the case the validation loss does not renew its minimum during 25 training epochs.

\begin{figure*}[t]
\centering
\includegraphics[width=0.8\linewidth]{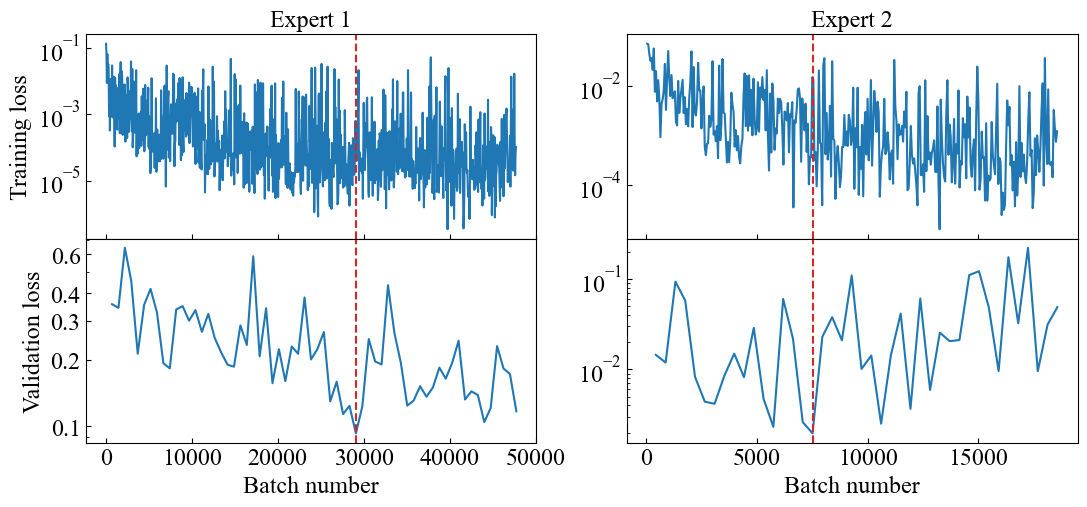}
\caption{
The training and validation loss curves for experts~1 (left) and 2 (right). The validation loss is estimated at the end of an epoch. The best checkpoints, corresponding to the minimum validation loss, are shown by the red dashed line.
}
\label{fig:train_val_loss}
\end{figure*}

Figure~\ref{fig:train_val_loss} shows the behavior of the training and validation loss curves for both models. As is seen from the figure, the training loss drops to low values almost immediately, and then continues to slightly decrease. Such a fast training speed is due to the relative simplicity of the data and prominent high-SNR FRB events in the training sample, which can be easily recognized by the neural network. The great level of noise in the curves is explained by the fact that the network is always nearby its good performance. 

The validation loss values for expert~1 (bottom left panel) are significantly higher than the training loss because the validation is performed on the ``hard'' events only (Table~\ref{tab:training}): the narrow and low-DM pulses. The validation curve is also noisy but clearly has a decreasing trend. The training was stopped after reaching the early stopping criterion (25 epochs without an improvement of validation loss).

The validation loss curve for expert~2 (bottom right panel) has an increasing trend at the right side, which shows that overfitting has begun. The best model has been chosen at the minimum of the validation loss, as well as for expert~1.

The number of epochs before reaching the minimum is 38 and 16 for experts~1 and 2, respectively. The training time corresponding to the curves shown in Fig.~\ref{fig:train_val_loss} is 85 and 33 min.

\section{Evaluation}
\label{sec:eval}

\subsection{The metrics used}

Due to the relative simplicity of the data and the presence of a large number of prominent synthetic FRB pulses in the sample, the classification metrics as measured from the test subsample are extremely high and the receiver operating characteristic curve is close to a rectangle in shape. For realistic evaluation of model performance, we found it is better to use instead (1)~the false positive (FP) rate, which is determined based on real noise records and RFI signals in the test sample, and (2)~the recall of FRB events with particular properties ($S/N$, $\rm DM$, and width):
\begin{equation}
\rm Recall = \frac{TP}{TP+FN},
\end{equation}
where $\rm TP$ and $\rm FN$ are the true positive and false negative detections. The recall, in other words, is the percentage of FRBs retrieved by the model.

\subsection{False positive rate}

The FP rate of the model output can be regulated by setting the probability threshold for detection. The neural network is trained in such a way that ${\rm logit} = 0$ at the output neuron corresponds to the 50\%\ probability estimate that an instance is an FRB event; the positive logits correspond to higher probabilities. For expert~1 we have left the zero threshold for event classification, which resulted in its FP rate of 0.65\%. As low as this value is, for the four radiometers it corresponds to about 2400~FP~events per day. For expert~2 the threshold ${\rm logit} = 0$ corresponds to an FP rate of 0.08\%. The resulting FP rate of the cascade is then $0.0065\times0.0008=0.0005\%$, or about 2 FP events per day.

\begin{table*}[t]
\small
\centering
\caption{False positive rate in the cascade scheme at different probability thresholds. 
The 95\%\ confidence intervals (CI) are estimated by 5000~bootstrap subsamples of the test sample
}
\label{tab:fp_rate}
\begin{tabular}{c||ccc|ccc||cc}
\hline
Probability & \multicolumn{3}{c|}{Expert~1} & \multicolumn{3}{c||}{Expert~2} & \multicolumn{2}{c}{Cascade} \\
threshold & logit thresh. & FP rate & 95\% CI & logit thresh. & FP rate & 95\% CI & FP rate & FP per day \\
\hline
low & $0$ & $0.0065$ &  $\pm 0.0008$ & $-4.1$ & 0.009 & $\pm 0.001$ & $5.9\times10^{-5}$ & 20 \\
medium & $0$ & $0.0065$ & $\pm 0.0008$ & $0$ & 0.0008 & $\pm 0.0003$ & $5.2\times10^{-6}$ & 2 \\
high & $0$ & $0.0065$ & $\pm 0.0008$ & $+4.1$ & $\sim 0$ & $\sim 0$ & $\sim 0$ & $\sim 0$ \\
\hline
\end{tabular}
\end{table*}

\begin{figure*}[t]
\centering
\includegraphics[width=0.32\linewidth]{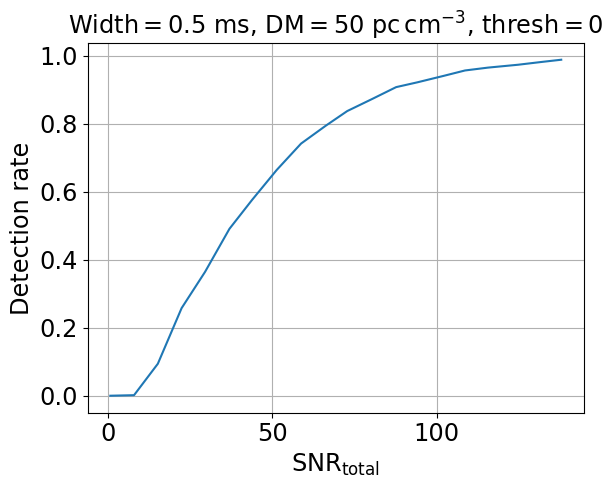}
\includegraphics[width=0.32\linewidth]{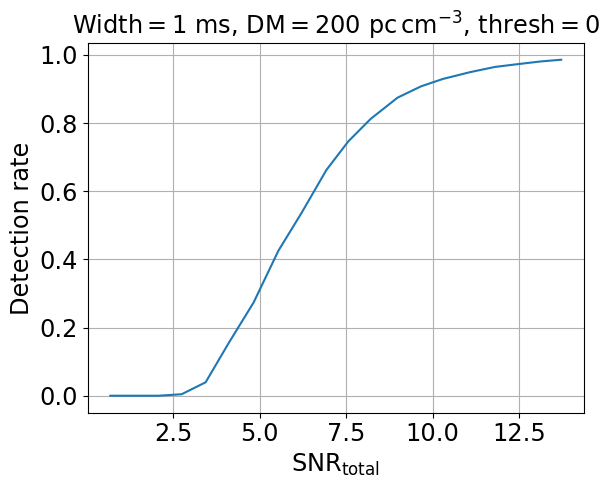}
\includegraphics[width=0.32\linewidth]{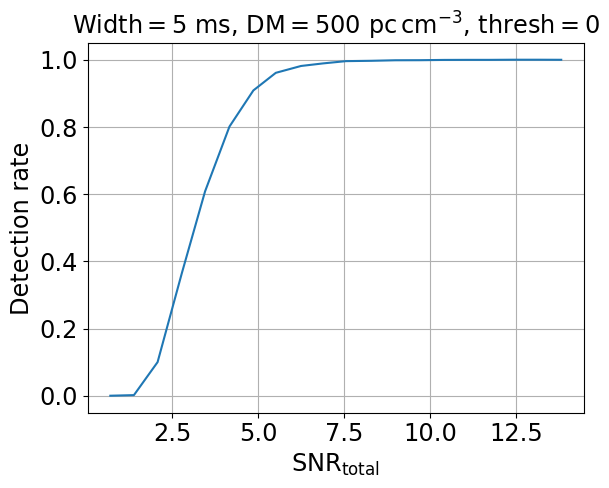}
\includegraphics[width=0.32\linewidth]{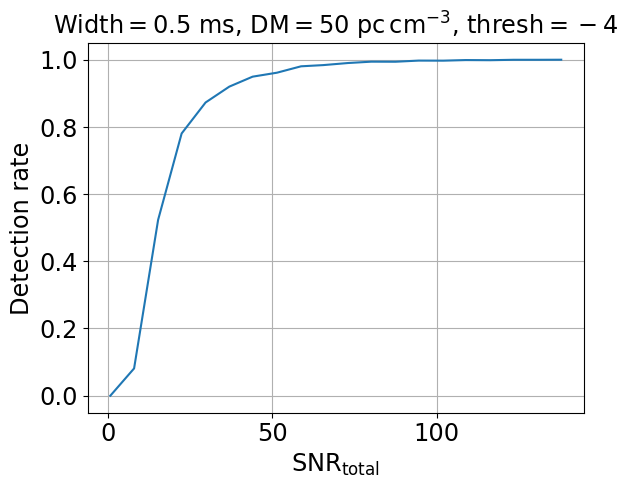}
\includegraphics[width=0.32\linewidth]{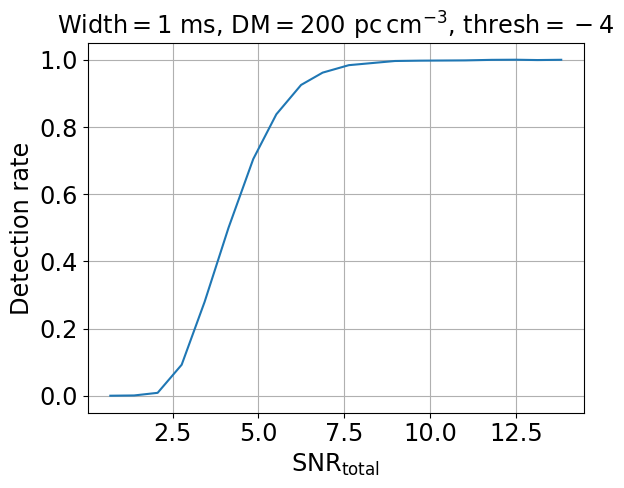}
\includegraphics[width=0.32\linewidth]{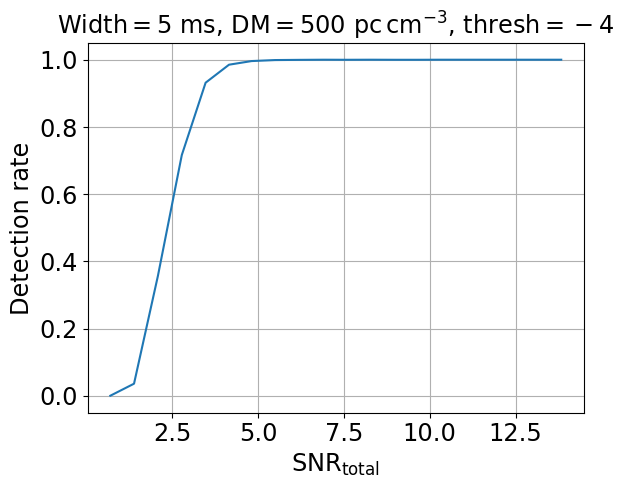}
\caption{
\label{fig:detection_rate} 
Detection rates in the cascade scheme depending on the total SNR. The panels (left to right) correspond to pulses with certain characteristics. The upper row corresponds to the medium (50\%, $\rm logit=0$) probability threshold, the lower row is for the low probability threshold ($\rm logit=-4.1$). 
The 95\%\ confidence intervals for the detection rates do not exceed 0.02.
}
\end{figure*}

By adjusting the probability threshold of expert~2 we can control the final FP rate (Table~\ref{tab:fp_rate}). For instance, setting the logit threshold to $+4.1$ resulted in the absence of FP detections in the test sample. We should note though that this also means that weaker FRB signals are going to be missed by the model if the threshold is set too high. We describe our strategy of setting different thresholds in Section~\ref{sec:inference}.

\subsection{FRB Recall}

The recall of synthetic FRB events varies depending on their characteristics, mainly on the dispersion measure and width. The lower these values, the harder it is for the network to distinguish an FRB from an RFI signal. This is demonstrated in Fig.~\ref{fig:detection_rate}, where we show the recall (detection rate) of FRB events for three particular set of characteristics ($w=0.5$~ms, $\rm DM=50$~pc\,cm$^{-3}$; $w=1$~ms, $\rm DM=200$~pc\,cm$^{-3}$; and $w=5$~ms, $\rm DM=500$~pc\,cm$^{-3}$) depending on the total signal-to-noise ratio of the signal, calculated here as 
\begin{equation}
\rm SNR_{total} = \sqrt{\sum_{ch} SNR_{ch}^2}, \qquad
\rm SNR_{ch} = \frac{\it A}{\sigma_{noise}},
\end{equation}
where $\rm SNR_{ch}$ is the signal-to-noise ratio in a channel, defined as the amplitude $A$ of a signal expressed in the noise variance $\sigma_{\rm noise}$.
To construct the curves in Fig.~\ref{fig:detection_rate}, we took 20 evenly spaced points on the SNR axis, calculated for each point 5000~synthetic pulses with given characteristics, and measured the detection rates by applying the cascade scheme. The 95\%\ confidence intervals for the detection rates have been measured by the bootstrap method and do not exceed 0.02. With such narrow confidence intervals, and the clear difference between the curves in Fig.~7, the significance tests (e.g., z-test for proportions) for the difference in detection rates for pulses with the same SNRs but different characteristics give extremely low p-values close to zero, i.e., the differences between the detection rate curves in Fig.~7 have very high significance.

Along with the better detection rate for broader pulses with greater DMs, a comparison of the upper and lower panels in the figure shows significant improvement in the detection rate with the lowered probability threshold (the output neuron $\rm logit=-4.1$). It should be remembered that this is achieved at the cost of higher FP rate (Table~\ref{tab:fp_rate}).

We can state that the model shows quite promising detection rates for ``typical'' FRBs with $w=5$~ms, \mbox{$\rm DM=500$}~pc\,cm$^{-3}$, while higher $\rm SNRs$ are required for more narrow pulses with lower $\rm DMs$. The lowered probability threshold is obviously better for detecting faint pulses.

\subsection{Validation on a giant pulse of the Crab Pulsar}

\begin{figure}[t]
\centering
\includegraphics[width=0.9\linewidth]{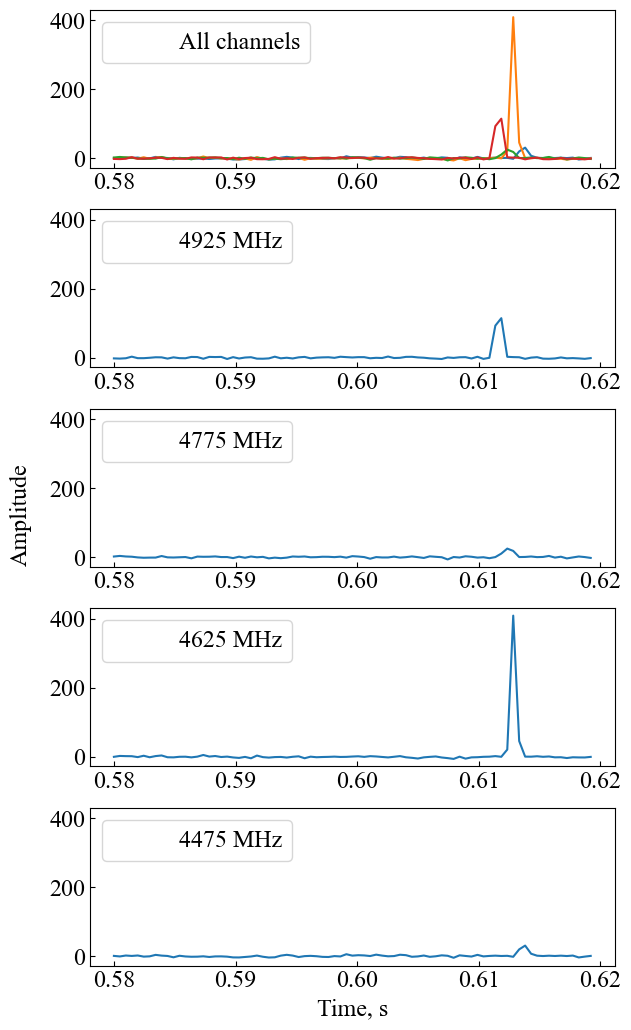}
\caption{
\label{fig:crab} 
A giant pulse from the Crab Pulsar observed on 30~June 2018.
}
\end{figure}

\begin{figure}[t]
\centering
\includegraphics[width=\linewidth]{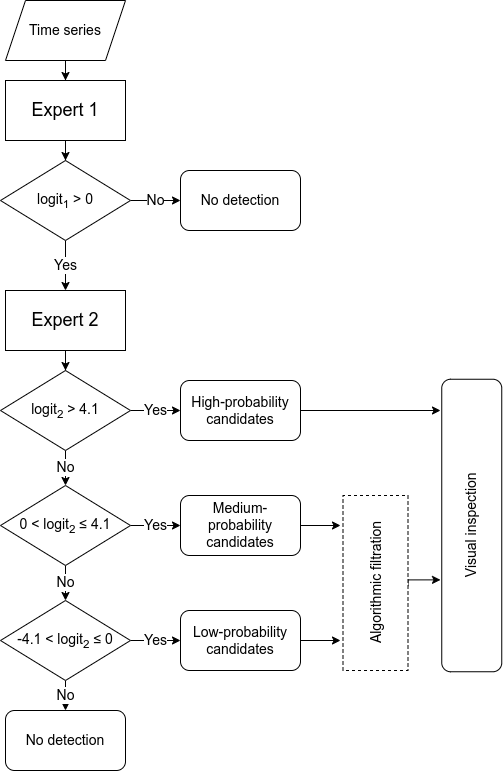}
\caption{
\label{fig:inference} 
A flowchart illustrating the inference workflow.
}
\end{figure}

As there are no FRB events detected with the specific RATAN-600 equipment so far, we tested the operability of the network using a giant pulse from the Crab Pulsar, which was observed on 30 June 2018 (Fig.~\ref{fig:crab}). The event demonstrates extremely narrow time profiles (the actual width is probably lower than the temporal resolution) and has a very low dispersion measure $\rm DM \sim 57$~pc\,cm$^{-3}$. This kind of a signal is not what is expected from FRB events, which are extragalactic sources with higher $\rm DM$ and broader time profiles, but it has a very high SNR, although with a clear scintillation pattern. The network has successfully detected the signal with the probability threshold of expert~2 set to $-4$. The detection of this pulse has also ensured that all the data preprocessing, such as the order of radio frequencies within the records, normalization, etc., works appropriately.

\section{Inference}
\label{sec:inference}
 
During inference, records from a radiometer, typically of 1-hour duration, are sliced into 1-s time intervals, which are fed into the cascade of the two neural networks. The records are beforehand reduced to a per-channel noise variance $\sigma_{\rm noise}=1$, as in the training sample. Each time interval is additionally checked for matching this noise level, which is for the case of possible location of the source in an extended radio emitting region, where noise variance can be greater than the average. 

We apply half-interval scanning, which means that each time interval is scanned twice, at different positions within the scanned image: e.g., 0--1, 0.5--1.5, \mbox{1--2}~s, etc. 

The flowchart illustrating the inference workflow is shown in Fig.~\ref{fig:inference}.
The time series is passed to expert~1 and then, in the case of detection (\mbox{$\rm logit>0$}), to expert~2. Based on the metrics obtained above during evaluation, we collect the positive signals into three groups: the high-probability (\mbox{$\rm logit_2>+4.1$}), medium-probability (\mbox{$\rm 0<logit_2\leq+4.1$}), and low-probability (\mbox{$\rm -4.1<logit_2\leq0$}) signals. The groups have different FP rates (Table~\ref{tab:fp_rate}), as is mentioned before.

We currently do not have another (algorithmic) mechanism to preselect or postselect candidate events, so the signals provided by the neural network cascade are to be examined visually. With a typical length of archive records of about 1~year, this manual post-processing is still somewhat a challenge for the groups of medium-probability and low-probability signals: as much as several thousands FP signals can pass the cascade during a year. For these cases we may apply additional two-stage filtration. Firstly, the radiometer has four beams, each oriented toward a different sky area, therefore an event detected at more than one radiometer is likely an FP signal, which can be omitted from consideration. The other filter is the demand for the signal to be detected in both records during half-interval scanning. With this filtration implemented, we reduce the amount of FP signals in a year to an order of a hundred, which is manageable for visual inspection, although implementation of a more sophisticated method of candidate selection might be necessary in the future.

\section{Conclusion}

Based on the EfficientNet family of models, we have developed a 1D convolutional neural network for processing multichannel time series and applied it to the problem of searching for FRB pulses in the RATAN-600 4-channel records. The input data for the network are time series with a duration of approximately 1~s (4080~elements with a time resolution of 0.245~ms).

The novelty of the presented EfficientNet1d-XS in terms of astronomical application is its operation with few-channel radio observations (time series). The small number of channels allows using 1D convolutions, which lowers the computational complexity by a factor of $k$, where $k$ is the convolution kernel size (with the typical value $k=3$). The network has a relatively small number of parameters ($\sim$1.13~mln), which allows operation on modest gaming GPUs. At the same time, the 19 MBConv layers, implemented in the EfficientNet family of models, provide the capacity to learn fine features in the data. The two-stage cascade scheme effectively suppresses false positive detections. The network can be trained to classify arbitrary few-channel time series, the code is available on GitHub.

The training dataset was collected using records of actual radiometer noise and observed radio frequency interference signals of various time profiles. The FRB events were synthetic, modeled as a Gaussian time profile modified further for scattering and dispersion smearing. We varied the synthetic pulses by amplitudes, widths, dispersion measures, spectral indices. A scintillation/nulling pattern was additionally applied to approximately 77\%\ of the synthetic pulses. The pulses we injected into actual records of radiometer noise. 

The trained neural network demonstrates a false positive detection rate less than 1\%, but this value is still too large, considering the number of instances to be analyzed and the absence of candidate preselection. To further reduce the number of false positive detections, we apply a cascade scheme with the second neural network (expert~2) trained on the false positive detections of the first network (expert~1). This effectively decreases false detections to the minimum: the FP rate of the cascade is a product of the FP rates of the two networks. The cascade scheme along with adjusting the probability threshold for expert~2 and a simple algorithmic filtration of signals from four radiometer beams with half-interval scanning allows us to suppress the FP detection rate 
to tolerable values suitable for final visual analysis.

The FRB detection rate is very high for the training dataset that has a large number of prominent synthetic events. Therefore we considered the true positive detection rate specifically for events with particular characteristics (width and DM). The analysis shows good performance for ``typical'' FRBs with $w=5$~ms and $\rm DM=500$~pc\,cm$^{-3}$: we expect to detect nearly all events with an SNR of the total pulse greater than 4. The detection rate, though, decreases for narrower pulses with smaller dispersion measures. This is a tradeoff between the false positive and true positive rates. For instance, the submillisecond pulses at very low DMs, characteristic of giant pulses from pulsars within our Galaxy, are detectable only at high SNRs. This is because the shape of such pulses and their location across radio channels are close to those of the pulse RFI, and it is harder for the network to distinguish between the two kinds. Improvement of the performance for narrower pulses is a matter of further development of either our cascade scheme or an algorithm for candidate pre/post-selection.

Other limitations that can be foreseen are as follows. The RFI in the training dataset are collected within a relatively short time interval of about a month. If some kinds of RFI signals have not been collected, the performance will degrade for these novel RFI types. Synthetic FRBs may not fully replicate real FRB characteristics, potentially affecting real-world performance. This is especially true for the FRB signals with complex non-gaussian time profiles, which are sometimes observed with other radio telescopes. As is already mentioned, the inference is still not fully automatic and rely on manual visual inspection at the final stage, which may be time-consuming on the scale of several years of the archived observations.

In this paper, we used a very general approach to set the neural network hyperparameters, such as the number of layers or the number of filters within a layer, and no real hyperparameter tuning has been performed. Because of the relative simplicity of the training data, the model quickly reaches low values of the loss function, close to the minimum. It might seem that the network architecture could have been simplified, but the comparison with the FCN shows that the complexity of EfficientNet has a positive effect on the detection rates of narrow pulses, while the training and inference times remain reasonable. At the same time, we do not see a reason to enlarge the complexity, as it is the good detection of narrow pulses by expert~1 that worsens its false positive rate and forces us to use the cascade scheme. In general, the hyperparameter tuning may somehow decrease the inference time, but we believe that further improvement should be an interplay of the network architecture, the cascade scheme, the range of the FRB parameters in the training sample, and other factors that could improve the result. For example, another approach to expert~2 operation can be used, such as inference from the 2D DM--time diagrams, or some kind of the focal loss function can be applied to better detect narrow pulses.

The evaluation of the network performance is based on synthetic FRB pulses and real RFI signals. The operability is checked using an observation of a giant pulse from the Crab Pulsar. The network is currently being applied to scan the archive records obtained on the \mbox{RATAN-600} Western Sector, some promising results have been obtained, yet to be analyzed. Currently there are no real FRBs to be presented and validate the performance on the example of real events. The neural network is trained based on general assumptions about FRB parameters: we tried to capture the complete range of possible widths, DMs, and amplitudes. This approach may turn out to be not optimal for the specific instrumental characteristics of the RATAN-600 radio telescope: most probably, the currently expected ranges are excessive, taking into account the high radio frequencies at which the radiometer operates, its sensitivity, small number of channels, etc. For instance, we may expect unrealistically high SNRs and DMs along with the lower ones. Detections of real FRBs would help estimate the real distribution of FRB parameters characteristic of the telescope, narrow the expected parameter ranges, and improve the performance by retraining the network based on the new assumptions.

\section*{CRediT authorship contribution statement}
{\bf D.\,O.~Kudryavtsev:} Writing~-- original draft, Methodology, Software, Formal analysis, Visualization.
{\bf S.\,A.~Trushkin:} Writing~-- review and editing, Resources.
{\bf P.\,G.~Tsybulev:} Methodology, Resources, Software.
{\bf V.\,A.~Stolyarov:} Methodology.

\section*{Declaration of competing interest}
The authors declare that they have no known competing financial interests or personal relationships that could have appeared to influence the work reported in this paper.

\section*{Acknowledgments}
We thank the referees for their valuable suggestions that helped improve the paper.
Observations with the SAO~RAS telescopes are supported by the Ministry of Science and Higher Education of the Russian Federation. The renovation of telescope equipment is currently provided within the national project ''Science and Universities.\!''

\section*{Data availability}
The code of the developed 1D convolutional neural network for multichannel time series analysis (based on \mbox{PyTorch}) and the code used to model synthetic FRB events are available on GitHub: \url{https://github.com/DKudryavtsev/EfficientNet1D_FRB}


\end{document}